\documentstyle[12pt]{article}

\newcommand \beq{\begin{eqnarray}}
\newcommand \eeq{\end{eqnarray}}
\newcommand{\bea}{\begin{eqnarray*}}
\newcommand{\eea}{\end{eqnarray*}}
\newcommand{\be}{\begin{equation}}
\newcommand{\ee}{\end{equation}}
\newcommand{\ba}{\begin{array}}
\newcommand{\ea}{\end{array}}
\newcommand{\bc}{\begin{center}}
\newcommand{\ec}{\end{center}}

\begin{document}

\title{
Random phase approximation and its extension for the quantum O(2) anharmonic
oscillator 
}
\author{
Zoheir Aouissat
\\
{Institut f\"ur Kernphysik, Technische Hochschule Darmstadt,}\\
{Schlossgarten 9, D-64289 Darmstadt, Germany \footnote{E-mail address : 
aouissat@crunch.ikp.physik.tu-darmstadt.de}    }
\\
C\'ecile Martin 
\\
{Groupe de Physique Th\'eorique, 
}\\
{  Institut de Physique Nucl\'eaire,} \\
{ F-91406 , Orsay Cedex, France \footnote{E-mail address :
    martinc@ipno.in2p3.fr} }
}

\date{}

\maketitle 
\vspace*{1cm}
\begin{abstract}
We apply the random phase approximation (RPA) and its extension called
renormalized RPA to the quantum anharmonic oscillator with an O(2) symmetry.
We first obtain the equation for the RPA frequencies in the standard and in
the renormalized RPA approximations using the equation of motion method. In
the case where the ground state has a broken symmetry, we check the
existence of a zero frequency in the standard and in the renormalized RPA
approximations. Then we use a time-dependent approach where the standard RPA
frequencies are obtained as small oscillations arround the static solution
in the time-dependent Hartree-Bogoliubov equation. We draw the parallel
between the two approaches. 
\end{abstract}


\newpage





 
 
 
 
   

 


\section{ Introduction}

 Our understanding of the low-energy behaviour of Quantum Chromodynamics
 calls for the developement of non-perturbative methods for quantum field theories.
 Some succes has been obtained by adapting to relativistic quantum fields 
 well-known non-perturbative methods used in the nuclear many-body problem.
 For instance, variational methods using Gaussian wave functionals (which
 are analogous to the Hartree- Fock-Bogoliubov (HFB) kind of approximations) 
 have been applied
 to self-interacting bosonic field theory \cite{1}, and to gauge field theory
 \cite{2}.   One important question, which is currently intensely debated in the 
 literature, is how do such non-perturbative
 approximations respect the symmetries of the theory. 
 
  In this regard, we would like in the the present work to revisit this important 
 question 
 within the framework of yet an other well known and equally successful 
 method from the many-body theory, the  
 random phase approximation (RPA) and its extension, 
 the renormalized RPA \cite{3,3a}. 
 The first of these two has experienced in the past quite a number of applications 
 in various fields either in condensed matter or the nuclear problem. The second one 
 on the other hand has, since its first formulation by Rowe \cite{7a}, 
 attracted very little interest till very recently. Beside its use in the conventional
 nuclear problem, there has been indeed  several attempts as of lately to extend 
 its application to quantum field theoretic models as well \cite{6,7}.     
  It is worth reminding that in the standard RPA approximation,
 two-body expectation values are evaluated in the  HFB mean-field vacuum.
 In the renormalized RPA approximation, however, these two-body expectation values
  are determined using the true RPA vacuum, inferring to the renormalized RPA 
 a self-consistent character. 
 
 Recently the  quasi-particle RPA approach  has been  applied to the
 linear sigma model in order to obtain a correct description of the global
 chiral symmetry broken phase. The standard mean-field  HFB approximation
 gives a finite mass to the pion. However, it has been shown that in the RPA
 approximation the pion is massless in accordance with Goldstone theorem
 \cite{4,5}.  A careful study of the finite temperature chiral phase transition 
 in this model  reveal, however,  that the HFB-RPA approximation leads to a 
 first order phase transition. This clearly is  
 an artefact of the approximation as it is well admitted that the chiral 
 transition in this model is of second order. This problem can be traced back to 
 the fact that the finite temperature induced transition does not happen in the RPA
 vacuum, but rather in the 
 self-consistently built ground-state, namely the HFB state.
 The latter is clearly a wrong vacuum for the theory as it doesn't possess a valley
 in the broken phase. The RPA fluctuations which were crucial in  correcting
 for these shortcomings \cite{4,5} are of no use in the present
 situation since these are implemented perturbatively.  
 Therefore, it is obvious that such an approach is bound to fail in describing
 the theory in the vicinity of the phase transition.    
 A possible solution to this problem might very well be in relaxing the quasi-boson 
 approximation
 inherent in the standard RPA approach. This  leads to a self-consistent version
 of the RPA very difficult to put in practice. Therefore, it is more convenient 
 to consider rather the renormalized RPA variant. This is what we propose to 
 study here.
 
 The renormalized RPA approximation has been considered in
 \cite{6} in the context of scalar $\lambda \phi^4$ field theory in $1+1$
 dimensions. In this work, the formulation of the RPA equations was based on
 the Dyson equation approach. A preliminary work applies the renormalized
 RPA to O(N) field theory \cite{7}. Here, we will focus on the
 two-dimensional quantum anharmonic oscillator with an O(2) symmetry. This
 purely quantum mechanical model allows to obtain analytical expressions
 without the problems of divergences occurring in quantum field theories. It
 shows the possibility to have a vacuum state with spontaneous broken
 symmetry. It has also the advantage that we can compare the results
 obtained with the standard and renormalized RPA approximations with exact
 numerical results. This numerical investigation shall be published in a
 future work. Here we will concentrate on the formal aspects of the theory. 
 A crucial point is to check the existence of a zero
 excitation frequency above the vacuum state with broken symmetry in our
 non-perturbative approximations ( this corresponds in quantum field theory
 to the Goldstone mode).  Since these quantum mechanical systems are used 
 for demonstrational purpose, we will disregard all problems related to the infrared
 divergences occurring due to the presence of this zero energy mode.
 
 The paper is organized as follows. In the first part, we derive 
 the renormalized RPA equations
 using the equation of motion method \cite{7a}. In the second part, we use a
 time-dependent approach. In this formalism, the standard RPA
 frequencies are obtained as small oscillations around the static solution
 in the time-dependent Hartree-Bogoliubov equation. Within this second
 approach, we are not able to go beyond the standard RPA approximation.
 However, it is interesting to  draw the parallel
 between the two approaches.


\section{Renormalized RPA equations from the equation of motion method}

\setcounter{equation}{0}


The Hamiltonian for the O(2) anharmonic oscillator reads : 
\be \label{1.1}
H= \frac{P_1^2}{2} +\frac{P_2^2}{2} + \frac{\mu}{2} [ X_1^2+(\tilde X_2+
<X_2>)^2] + g [X_1^2 +(\tilde X_2+< X_2>)^2]^2 +
\eta (\tilde X_2+< X_2>) \ , \ee
where we have considered an explicit ($\eta \ne 0$) symmetry breaking and a
spontaneous ($<X_2> \ne 0$) symmetry breaking along the $X_2$
direction. In the $\mu <0$ case, the potential has a ``Mexican hat'' shape. 

The case of two particles in a harmonic potential and coupled by a linear
interaction, which is analytically solvable, has been considered by the
authors of reference \cite{7b} to demonstrate that the RPA correlation
formula works well. 

Let us define creation and annihilation operators $a_i^{+} $ and $a_i, i=1,2$,
according to : 
\be \label{1.2}
X_1=\frac{1}{\sqrt{2 \omega}} (a_1 +a_1^{+}) \quad \quad P_1=i \sqrt{\frac{\omega}{2}}
(a_1^{+}-a_1) \ , \ee 
\be \label{1.3}
\tilde X_2=\frac{1}{\sqrt{2 \Omega}} (a_2 +a_2^{+}) 
\quad \quad P_2=i \sqrt{\frac{\Omega}{2}}
(a_2^{+}-a_2) \ . \ee
The transverse frequency $\omega$, the radial frequency $\Omega$ and the
condensate $< X_2>$ will be determined self-consistently. Using the analogy
with the linear sigma model in quantum field theory, the $X_1$ and $X_2$
modes represent the pion and the sigma fields respectively. In terms of the
operators $a_i$ and $a_i^{+}$, the Hamiltonian reads : 
\be \ba{lll} \label{1.4}
H=
& {\displaystyle \sum_{i=1,2} p_{ii} a_i^{+}a_i + 
  \sum_{i=1,2} p_{i0} (a_i^{+}a_i^{+} + a_i
a_i)   } \\
& {\displaystyle + g_{11} (a_1^{+} + a_1)^4 + g_{12} (a_1^{+}+a_1)^2
  (a_2^{+}+a_2)^2 + g_{22} (a_2^{+} + a_2) ^4  
  } \\
& {\displaystyle + h_{12} (a_1^{+}+a_1)^2 (a_2^{+}+a_2) +h_{22}
  (a_2^{+}+a_2)^3 + \tilde \eta (a_2^{+}+a_2) + C 
} \ea \ , \ee 
where C is the following constant 
\be \label{1.5} 
C= \frac{\omega}{4} +\frac{\Omega}{4} +\frac{\mu}{4 \omega}+ 
\frac{\mu}{4 \Omega} + \frac{\mu}{2}< X_2>^2 + \frac{3g}{\Omega}
< X_2>^2 +  \frac{g}{\omega}< X_2>^2 + g < X_2>^4 + \eta
< X_2> \ . \ee 
We have defined the following quantities : 
\be \label {1.6}
p_{11}= \frac{\omega}{2} +\frac{\mu}{2 \omega} +\frac{2g}{\omega} <
X_2>^2 \ , 
\ee
\be \label {1.7}
p_{22}= \frac{\Omega}{2} +\frac{\mu}{2 \Omega} +\frac{6g}{\omega} 
<X_2>^2 \ , 
\ee
\be \label{1.8} 
p_{10}=\frac{1}{2}(p_{11}-\omega) \: \: \ , \: \: 
p_{20}=\frac{1}{2}(p_{22}-\Omega) \ , 
\ee
\be \label{1.9} 
g_{11}=\frac{g}{4 \omega^2} \: \: \ , \: \: g_{12}=\frac{g}{2  \omega \Omega} 
\: \: \ , \: \: g_{22}=\frac{g}{4 \Omega^2} \ , \ee
\be \label{1.10} 
h_{12}=\frac{2g}{\omega} \frac{1}{\sqrt{2 \Omega}} < X_2> \: \: \ , \:
\: h_{22}=\frac{2g}{\Omega} \frac{1}{\sqrt{2 \Omega}} < X_2>  \ , \ee
\be \label{1.11} 
\tilde \eta= \frac{1}{\sqrt{2 \Omega}} \left( \eta + \mu < X_2> + 4 g
< X_2>^3 \right) \ . \ee

\subsection{Mean Field Equations}

Normal ordering with respect to $a_i$ and $a^{+}_i$ gives the result : 
\be \label{1.12} \ba{lllllll}
H= & {\displaystyle \sum_{i=1,2} p_{ii} a_{i}^{+} a_i + \sum_{i=1,2} p_{i0}
  (a^{+}_ia^{+}_i +a_i a_i) } \\ 
& {\displaystyle + g_{11} \left \{ a_1^{+}{}^4 + a_1^4 + 6
    a^{+}_1a_1^{+}a_1a_1 + 4  a^{+}_1a_1^{+}a^{+}_1a_1 + 4  a^{+}_1a_1a_1a_1
    + 12 a^{+}_1 a_1 + 6 a^{+}_1a^{+}_1 +6 a_1 a_1 \right \} } \\
& {\displaystyle + g_{22} \left \{ a_2^{+}{}^4 + a_2^4 + 6
    a^{+}_2a_2^{+}a_2a_2 + 4  a^{+}_2a_2^{+}a^{+}_2a_2 + 4  a^{+}_2a_2a_2a_2
    + 12 a^{+}_2 a_2 + 6 a^{+}_2a^{+}_2 +6 a_2 a_2 \right \} } \\ 
& {\displaystyle + g_{12} \left \{ a_1^{+}{}^2 a_2^{+}{}^2 + a_1^{+}{}^2
    a_2^2 +a_1^2 a_2^{+}{}^2+a_1^2a_2^2 + 2 a_1^{+}{}^2 a_2^{+}a_2 + 2 a_1^2
    a_2^{+}a_2 \right. } \\
& {\displaystyle \left. 
  + 2 a_1^{+}a_1 a_2^{+}{}^2 + 4 a_1^{+}a_1 a_2^{+} a_2 +
  a_1^{+}{}^2 + a_1^2 + 2 a_1^{+}a_1 + a_2^{+}{}^2 + a_2^2 + 2 a_2^{+} a_2
  \right \} } \\ 
& {\displaystyle + h_{12} \left \{ a_1^{+}{}^2 a_2^{+} + a_1^{+}{}^2a_2 +
    a_1^2 a_2^{+} + a_1^2 a_2 + 2 a_1^{+}a_1 a_2^{+} + 2 a_1^{+}a_1 a_2
  \right \} } \\ 
& {\displaystyle + (h_{12}+3 h_{22} + \tilde \eta) (a_2 + a_2^{+}) + E_{HFB } }
\ea \ , \ee
where 
\be \ba{ll} \label{1.13}
E_{HFB}= & {\displaystyle
  \frac{\omega}{4}+\frac{\Omega}{4}+\frac{\mu}{4 \omega}+\frac{\mu}{4 \Omega}
+ \left( \frac{\mu}{2}+\frac{3g}{\Omega}+ \frac{g}{\omega}+
g< X_2>^2 \right) 
< X_2>^2 } \\
& {\displaystyle +\frac{3g}{4 \omega^2}+\frac{3g}{4 \Omega^2} + \frac{g}{2
    \omega \Omega} + \eta < \tilde X_2> } \ea \ee
is the mean field energy. 

Minimization of $E_{HFB}$ with respect to $< X_2>$ gives the equation for
the condensate  : 
\be \label{1.14}
\tilde \eta + 3 h_{22} + h_{12} =0 \ee 
that is 
\be \label{1.15} 
\eta + < X_2> \left[ \mu + 4 g <  X_2>^2 +
\frac{6g}{\Omega}+\frac{2g}{\omega} \right] \ . \ee
Minimization of $E_{HFB}$ with respect to $\omega$ and $\Omega$ gives the two
gap equations : 
\be \label{1.16} 
\omega^2 = \mu +4 g < X_2>^2 + \frac{6g}{\omega} + \frac{2g}{\Omega} 
\ , \ee
\be \label{1.17} 
\Omega^2 = \mu +12 g < X_2>^2 + \frac{2g}{\omega} + \frac{6g}{\Omega} \
. \ee
We check that the coefficient of the linear term in the expression
(\ref{1.12}) gives the equation for the condensate (\ref{1.14}). 

Equations (\ref{1.15})-(\ref{1.17}) co\"\i ncide with those written by
Stevenson in \cite{8}. When there is no explicit symmetry breaking
($\eta=0$), we have two solutions : one with $< X_2>=0$
and one where the symmetry is spontaneously broken $< X_2> \ne 0$. 
For the symmetric solution,  we have $\omega=\Omega$ and the
gap equation is 
\be \label{1.17a}
\omega^3-\mu \omega-8g=0 \ . \ee
We can compare the mean field
results with exact numerical calculations \cite{9}. For instance for $\mu=1$
and $g=1$, the energies of the ground state and the two first excited states
are in the mean-field approximation : $E_0=1,74015, E_1=3,90645$ and
$E_2=6,07275$ whereas exact numerical 
calculations give : $E_0=1,7242, E_1=3,8304$ and
$E_2=6,214$. However, we are not
aware of numerical calculations for the two-dimensional anharmonic
oscillator in the case $< X_2> \ne 0$. 

\subsection{RPA equations from the equation of motion method}

The symmetry generator, {\it i.e.} the angular momentum operator around the
3-axis, is given by : 
\be \label{1.18}
L_3=X_1 P_2 - (\tilde X_2+< X_2>) P_1 \ , 
\ee
or in terms of the creation and annihilation operators : 
\be \label{1.19}
L_3=\frac{i}{2} (\sqrt{\frac{\Omega}{\omega}} + \sqrt{ \frac{\omega}{\Omega}}) 
(a_1 a_2^{+}-a_1^{+}a_2) -\frac{i}{2} (\sqrt{\frac{\Omega}{\omega}} -
\sqrt{\frac{\omega}{\Omega}}) (a_1a_2-a_1^{+}a_2^{+})
-i\sqrt{\frac{\omega}{2}} < X_2> (a_1^{+}-a_1) \ . \ee

To derive the RPA equations, we will first use the equation of motion method
due to Rowe \cite{7a}. We assume that an exact eigenstate $\vert \nu>$ of
the Hamiltonian can be created from the exact vacuum $\vert 0>$ by 
an excitation operator $Q_{\nu}^{+}$ : 
\be 
\vert \nu > =Q_{\nu}^{+} \vert 0> \quad\hbox{and}\quad Q_{\nu}\vert 0>=0 \ee
Minimization of the energy $E_{\nu}=<\nu \vert H \vert \nu>/<\nu \vert \nu
>$ with respect to a variation $\delta Q_{\nu}$ of the operator $Q_{\nu}$
leads to the following set of equations : 
\be \label{1.19b}
<0 \vert [\delta Q_{\nu},[H,Q_{\nu}^{+}]] \vert 0>=\Omega_{\nu} <0\vert
[\delta Q_{\nu}, Q_{\nu}^{+}] \vert 0> \ , \ee
where $\Omega_{\nu}=E_{\nu}-E_0$ is the excitation energy. 
One has also the supplementary condition : 
\be \label{1.20}
<0 \vert [H,Q_{\nu}] \vert 0>=0 \ , \ee
which is equivalent to  generalized mean field equations \cite{7c}. 

We will restrict our choice of excitation operators $Q_{\nu}^{+}$ to those
which contain the same operators that appear in the symmetry generator
(\ref{1.19}), that
is : 
\be \label{1.21}
Q_{\nu}^{+}=U_{\nu}^{(1)} a_1^{+}-V_{\nu}^{(1)} a_1 + 
U_{\nu}^{(2)} a_1^{+} a_2^{+}-V_{\nu}^{(2)} a_1 a_2 + 
U_{\nu}^{(3)} a_1 a_2^{+}-V_{\nu}^{(3)} a_1^{+} a_2 \ . \ee 
The RPA ground state is defined by 
\be \label{1.22}
Q_{\nu} \vert RPA >=0 \ , \ee
and the expectation values appearing in eqs. (\ref{1.19}) and (\ref{1.20})
are taken in the RPA ground state.
RPA equations (\ref{1.19}) become matrix equations which allow to determine
the excitation energy $\Omega_{\nu}$ and the amplitudes $U_{\nu}$ and
$V_{\nu}$ : 
\be \label{1.23} 
\pmatrix{ {\cal A} & {\cal B} \cr -{\cal B} & -{\cal A} \cr } \:   
\pmatrix{ {\cal U}_{\nu} \cr {\cal V}_{\nu} \cr } = \Omega_{\nu} 
 \: {\cal N} \: 
\pmatrix{ {\cal U}_{\nu} \cr {\cal V}_{\nu} \cr }  
\ , \ee 
where we have defined the three-dimensional vectors : 
\be \label{1.24}
{\cal U}_{\nu}=\pmatrix{ U_{\nu}^{(1)} \cr U_{\nu}^{(2)} \cr U_{\nu}^{(3)}
  \cr }  \quad \ , \quad
{\cal V}_{\nu}=\pmatrix{ V_{\nu}^{(1)} \cr V_{\nu}^{(2)} \cr V_{\nu}^{(3)}
  \cr } \ . \ee
The norm matrix is given by : 
\be \label{1.24a} 
{\cal N}=\pmatrix{ {\cal N}_A & {\cal N}_B \cr {\cal N}_B & {\cal N}_A \cr }
\ , \ee
where the $3 \times 3$ matrices ${\cal N}_A$ and ${\cal N}_B$ are :  
\be \label{1.25}
{\cal N}_A=\pmatrix{ 1 & 0 & 0 \cr 0 & 1+\tau_1+\tau_2 & \kappa_1 \cr 
0 & \kappa_1 & \tau_1-\tau_2 \cr } \quad \ , \quad   
{\cal N}_B=\pmatrix{ 0 & 0 & 0 \cr 0 & 0 & -\kappa_2 \cr 
0 & \kappa_2 & 0 \cr } \ . \ee 
We have introduced the notations : 
\be \label{1.26}
\kappa_1=<a_1^{+}a_1^{+}>=<a_1 a_1> \quad \ , \quad 
\kappa_2=<a_2^{+}a_2^{+}>=<a_2 a_2> \ee
\be \label{1.27} 
\tau_1=<a_1^{+}a_1> \quad \ , \quad \tau_2=<a_2^{+}a_2> \ee
We have assumed that the previous quantities are real.
We have also used by definition $<a_2^{+}>=<a_2>=0$. 

The RPA basis being complete, we have the following expression for any
operator ${\cal O}$ in terms of the RPA excitation operators $Q_{\nu}^{+}$
and $Q_{\nu}$ : 
\be \label{1.26a} 
{\cal O}=\sum_{\nu=1,3} Q_{\nu}^{+} \left< [Q_{\nu},{\cal O}]\right>-Q_{\nu} 
\left< [Q_{\nu}^{+},{\cal O}] \right> \ , \ee
where the expectation values are taken on the RPA ground state. We have
therefore the expression of the operators $a_1^{+}, a_1^{+} a_2^{+}$ and
$a_1^{+}a_2$ in terms of $Q_{\nu}^{+}$ and $Q_{\nu}$ and we deduce : 
\be \label{1.28} 
<a_1^{+}>=<a_1^{+}a_2^{+}>=<a_1^{+}a_2>=0 \ . \ee
We will introduce the following quantities : 
\be \label{1.29}
\Gamma = \tau_1 + \kappa_1 +\tau_2 + \kappa_2 + 1 
\ , \ee 
\be \label{1.30}
\Delta=\tau_1+\kappa_1 - \tau_2 -\kappa_2 \ . \ee

An important condition to remember in the EOM approach for RPA equations is
that the norm matrix ${\cal N}$ has to be invertible in order to be able to
write the normalization condition for the excited state $\vert \nu>$.  

The matrix elements of ${\cal A}$ and ${\cal B}$ are defined according to : 
\be 
{\cal A}_{11}=<[a_1,[H,a_1^{+}]]> \ee
\be 
{\cal A}_{12}=<[a_1,[H,a_1^{+}a_2^{+}]]> \ee
\be 
{\cal A}_{13}=<[a_1,[H,a_1a_2^{+}]]> \ee
\be 
{\cal B}_{11}=-<[a_1,[H,a_1]]> \ee
\be 
{\cal B}_{12}=-<[a_1,[H,a_1a_2]]> \ee
\be 
{\cal B}_{13}=-<[a_1,[H,a_1^{+}a_2]]> \ee

\be 
{\cal A}_{21}=<[a_1 a_2,[H,a_1^{+}]]> \ee
\be 
{\cal A}_{22}=<[a_1 a_2,[H,a_1^{+}a_2^{+}]]> \ee
\be 
{\cal A}_{23}=<[a_1 a_2,[H,a_1 a_2^{+}]]> \ee
\be 
{\cal B}_{21}=-<[a_1 a_2,[H,a_1]]> \ee
\be 
{\cal B}_{22}=-<[a_1 a_2,[H,a_1a_2]]> \ee
\be 
{\cal B}_{23}=-<[a_1 a_2,[H,a_1^{+}a_2]]> \ee

\be 
{\cal A}_{31}=<[a_1^{+} a_2,[H,a_1^{+}]]> \ee
\be 
{\cal A}_{32}=<[a_1^{+} a_2,[H,a_1^{+}a_2^{+}]]> \ee
\be 
{\cal A}_{33}=<[a_1^{+} a_2,[H,a_1a_2^{+}]]> \ee
\be 
{\cal B}_{31}=-<[a_1^{+} a_2,[H,a_1]]> \ee
\be 
{\cal B}_{32}=-<[a_1^{+} a_2,[H,a_1a_2]]> \ee
\be 
{\cal B}_{33}=-<[a_1^{+} a_2,[H,a_1^{+}a_2]]> \ee

We need to calculate all these commutators. In the renormalized RPA
approximation, when calculating the expectation values in the RPA ground
state, one uses the following approximation : 
\be \label {1.31} 
<\alpha_i \alpha_j \alpha_k \alpha_l> \simeq <\alpha_i \alpha_j> <\alpha_k
\alpha_l> +<\alpha_i \alpha_k> <\alpha_j \alpha_l> + <\alpha_i \alpha_l>
<\alpha_j \alpha_k> \ , \ee
where the $\alpha$ operators are either creation or annihilation operators
of the bosons 1 and 2. The expectation values are calculated self-consistently
whereas in standard RPA, they are calculated in the HFB ground state.
We obtain the following expressions for the matrix elements of ${\cal A}$
and ${\cal B}$ : 
\be 
{\cal A}_{11} \equiv {\cal E}_{\pi}=p_{11} +12 g_{11} (\Gamma + \Delta) +
2g_{12} (\Gamma-\Delta) \ee
\be 
{\cal A}_{12}=\tilde \eta +3h_{22}(\Gamma-\Delta)+h_{12}(\Delta+3 \Gamma) \ee
\be
{\cal A}_{13}=2 h_{12} \Delta \ee
\be 
{\cal A}_{21}=2 h_{12} \Gamma \ee
\be 
{\cal A}_{22}=({\cal E}_{\pi}+{\cal E}_{\sigma}) (\tau_1+\tau_2+1)
+\kappa_1 \chi_{\pi} + \kappa_2 \chi_{\sigma} + 4g_{12} \Gamma^2 \ee
\be 
{\cal A}_{23}=-\kappa_1 ({\cal E}_{\pi}-{\cal E}_{\sigma}) -(1+\tau_1
+\tau_2) \chi_{\pi} + 4 g_{12} \Gamma \Delta \ee 
\be 
{\cal A}_{31}=2 h_{12} \Delta \ee 
\be 
{\cal A}_{32}=\kappa_1 ({\cal E}_{\pi} + {\cal E}_{\sigma}) +
(\tau_1-\tau_2) \chi_{\pi} + 4 g_{12} \Gamma \Delta \ee 
\be 
{\cal A}_{33}=(\tau_2-\tau_1) ( {\cal E}_{\pi}-{\cal E}_{\sigma}) -\kappa_1
\chi_{\pi} -\kappa_2 \chi_{\sigma} + 4 g_{12} \Delta^2 \ee

\be 
{\cal B}_{11} \equiv \chi_{\pi}=2 p_{10}+2g_{11}(\Gamma + \Delta) + 2 g_{12}
(\Gamma -\Delta) \ee 
\be 
{\cal B}_{12}=2 h_{12} \Gamma \ee 
\be 
{\cal B}_{13}=\tilde \eta + h_{12} (\Gamma + 3 \Delta) + 3 h_{22}
(\Gamma-\Delta) \ee 
\be 
{\cal B}_{21}=2 h_{12} \Gamma \ee 
\be 
{\cal B}_{22}=\kappa_1 \chi_{\sigma} + \kappa_2 \chi_{\pi} + 4 g_{12}
\Gamma^2 \ee 
\be 
{\cal B}_{23} = -\kappa_2 ( {\cal E}_{\pi} -{\cal E}_{\sigma}) + (1 + \tau_1
+ \tau_2) \chi_{\sigma} + 4 g_{12} \Gamma \Delta \ee
\be 
{\cal B}_{31}=2 h_{12} \Delta \ee
\be 
{\cal B}_{32} = -\kappa_2 ({\cal E}_{\pi}+ {\cal E}_{\sigma}) + (\tau_1
-\tau_2) \chi_{\sigma} + 4 g_{12} \Gamma \Delta \ee
\be 
{\cal B}_{33}=\kappa_2 \chi_{\pi} + \kappa_1 \chi_{\sigma} + 4 g_{12}
\Delta^2 \ee

The quantities ${\cal E}_{\pi}, {\cal E}_{\sigma}, \chi_{\pi}$ and
$\chi_{\sigma}$ are defined according to : 
\be \label{1.32} 
{\cal E}_{\pi}=<[a_1,[H,a_1^{+}]]> \ee
\be \label{1.33} 
{\cal E}_{\sigma}=<[a_2,[H,a_2^{+}]]> \ee
\be \label{1.34} 
\chi_{\pi}=-<[a_1,[H,a_1]]> \ee
\be \label{1.35} 
\chi_{\sigma}=-<[a_2,[H,a_2]]> \ee

The equations for $< X_2>$ and the frequencies $\omega$ and $\Omega$
(or for ${\cal E}_{\pi}$ and ${\cal E}_{\sigma}$) are respectively obtained
by writing the generalized mean-field equations (\ref{1.20}) : 
\be \label{1.36} 
<[H,a_2]>=0 \ee 
\be \label{1.37} 
<[H,a_1^{+} a_1^{+}]>=0 \ee
\be \label{1.38}
<[H,a_2^{+} a_2^{+}]>=0 \ee
They are equivalent to the minimization of the generalized mean-field
energy $<H>$ with respect to $< X_2>$, $\omega$ and $\Omega$ ( see
the expression of $<H>$ eq.(\ref{1.77}) below). We obtain : 
\be \label{1.39} 
\tilde \eta +3 h_{22}(\Gamma -\Delta) + h_{12} (\Gamma +\Delta) =0 \ee 
\be \label{1.40} 
p_{11} \kappa_1 + p_{10} (2 \tau_1 +1) + 6 g_{11} (\Gamma+ \Delta)^2 + g_{12}
(\Gamma + \Delta) (\Gamma-\Delta)=0 \ee
\be \label{1.41}
p_{22} \kappa_2 + p_{20} (2 \tau_2 +1) + g_{12}
(\Gamma + \Delta) (\Gamma-\Delta)+ 6 g_{22} (\Gamma- \Delta)^2 =0 \ee
By using the definitions of $p_{10},p_{20}, {\cal E}_{\pi}$ and ${\cal
  E}_{\sigma}$, the two last equations, which we call the generalized 
gap equations, can be written as : 
\be \label{1.42}
{\cal E}_{\pi} (\Gamma +\Delta) = \omega (2 \tau_1 +1) \ee 
\be \label{1.43}
{\cal E}_{\sigma} (\Gamma -\Delta) = \Omega (2 \tau_2 +1)  \ee

When using the generalized mean-field equations (\ref{1.39}), (\ref{1.42})
and (\ref{1.43}), we check that the matrices 
${\cal A}$ and ${\cal B}$ are symmetric. 

\subsection{Standard RPA approximation} 

In the standard RPA approximation, all expectation values are taken in the
HFB ground state. We  therefore have : $\tau_1=\tau_2=\kappa_1=\kappa_2=0$ and
$\Gamma=1, \Delta=0$.  The generalized mean-field equations (\ref{1.39}), 
(\ref{1.42})
and (\ref{1.43}) reduce to the mean-field equations given in subsection
(2.1) , eqs. (\ref{1.15}), (\ref{1.16}) and (\ref{1.17}). We have : ${\cal
  E}_{\pi}=\omega, {\cal E}_{\sigma}=\Omega, \chi_{\pi}=\chi_{\sigma}=0$.
The matrices ${\cal A}$ and ${\cal B}$ become much simpler : 
\be \label{1.44}
{\cal A}=\pmatrix{ \omega & 2 h_{12} & 0 \cr 2 h_{12} & \omega +\Omega + 4
  g_{12} & 0 \cr 0 & 0 & 0 \cr } \quad \ , \quad
{\cal B}=\pmatrix{ 0 & 2h_{12} & 0 \cr 2 h_{12} & 4 g_{12} & 0 \cr 0 & 0 & 0
  \cr} \ , \ee
where we have used the equation for $< X_2>$ : $\tilde \eta +3h_{22}+
h_{12}=0$. 
We note that 
\be \label{1.45}
{\cal A}= \pmatrix{ \omega & 0 & 0 \cr 0 & \omega + \Omega & 0 \cr
0 & 0 & 0 \cr } + {\cal B} \ . \ee
The matrix ${\cal N}$ is diagonal  : 
\be \label{1.46} 
{\cal N}_A=\pmatrix{ 1 & 0 & 0 \cr 0 & 1 & 0 \cr 0 & 0 & 0 \cr} \quad \ ,
\quad {\cal N}_B=\pmatrix{ 0 & 0 & 0 \cr 0 & 0 & 0 \cr 0 & 0 & 0 \cr} \ . \ee
We see that the amplitudes $U_{\nu}^{(3)}$ and $V_{\nu}^{(3)}$ respectively
in front of the operators $a_1 a_2^{+}$ and $a_1^{+} a_2$ in the excitation
operator $Q_{\nu}^{+}$ (\ref{1.21}) decouple. In the symmetric phase, we have
$L_3=i(a_1 a_2^{+}-a_1^{+} a_2)$ and the operators which decouple are those
which appear in the symmetry generator. 

The standard RPA equations therefore reduce to a $4 \times 4$ system
(instead of 6 by 6). Its dimension will be then divided by two and the
RPA frequencies $\Omega_{\nu}$ satisfy : 
\be \label{1.47}
\det \left[ \left( {\cal A}-{\cal B} \right) \left( {\cal A} + {\cal B}
\right) - \Omega_{\nu} I \right] =0 \ , \ee
where now ${\cal A}$ and ${\cal B}$ are 2 by 2 matrices and I is the 2 by 2
unity matrix. This gives : 
\be \label{1.48} 
\left( \omega^2 - \Omega_{\nu}^2 \right) \left( (\omega + \Omega) -
  \Omega_{\nu}^2 + 8 g_{12}(\omega+\Omega) \right) -16 h_{12}^2 \omega
  (\omega+\Omega) =0 \ . \ee
This equation is valid for the two solutions : $< X_2>=0$ and $<
X_2> \ne 0$. 

Let us check the existence of a zero RPA frequency in the case of a ground
state with broken symmetry (the analog of the Goldstone mode in the linear
sigma model). (In the following, we have no explicit symmetry breaking :
$\eta=0$).  
By introducing the two following quantities (which correspond to loop
integrals in quantum field theories) : 
\be \label{1.49}
I_{\pi}=\frac{1}{2 \omega} \quad  \ , \quad I_{\sigma}=\frac{1}{2 \Omega} \
, \ee
the two gap equations in the case $< X_2> \ne 0$  can be written as : 
\be \label{1.50} 
\omega^2=8 g (I_{\pi}-I_{\sigma}) \ , \ee 
\be \label{1.51} 
\Omega^2=8 g < X_2>^2 \ . \ee 
We then introduce the following quantity (which is analog to the self-energy
in quantum field theories) : 
\be \label{1.52}
\Sigma (\Omega_{\nu}^2)=\frac{\omega + \Omega}{2 \omega
  \Omega}\frac{1}{\Omega_{\nu}^2-(\omega+ \Omega)^2} \ . \ee 
Equation (\ref{1.48}) for the RPA frequencies can be rewritten as : 
\be \label{1.53}
\left( \Omega_{\nu}^2 -\omega^2 \right) \left( 1 -4g 
  \frac{\omega + \Omega}{ \omega
  \Omega}\frac{1}{\Omega_{\nu}^2-(\omega+ \Omega)^2} \right) -32 g^2 <
X_2>^2 \frac{\omega + \Omega}{ \omega
  \Omega}\frac{1}{\Omega_{\nu}^2-(\omega+ \Omega)^2} =0 \ , \ee 
or 
\be \label{1.54} 
\Omega_{\nu}^2-\omega^2=64 g^2 < X_2>^2 \frac{\Sigma(\Omega_{\nu}^2)}{1
  - 8 g \Sigma(\Omega_{\nu}^2)} \ . \ee
This equation is identical to eq. (28) of \cite{6}. 
We also notice that : 
\be \label{1.55} 
\Sigma(\Omega_{\nu}=0)=-\frac{1}{2 \omega \Omega (\omega + \Omega)} \ , \ee 
\be \label{1.56} 
I_{\pi}-I_{\sigma}=\frac{\Omega^2-\omega^2}{2 \omega \Omega (\omega +
  \Omega)} \ . \ee 
Therefore : 
\be \label{1.57} 
\Sigma(\Omega_{\nu}=0)=\frac{I_{\pi}-I_{\sigma}}{\omega^2 -\Omega^2} \ . \ee 
We then use the gap equation (\ref{1.51}) to write the equation for the RPA
frequencies (\ref{1.54}) as : 
\be \label{1.58} 
\Omega_{\nu}^2 =\omega^2 + 8 g \Omega^2 \frac{\Sigma(\Omega_{\nu}^2)}{1
  - 8 g \Sigma(\Omega_{\nu}^2)} \ ,  \ee  
and then we use the first gap equation (\ref{1.50}) to obtain : 
\be \label{1.59}
\Omega_{\nu}^2=\frac{8g}{1
  - 8 g \Sigma(\Omega_{\nu}^2)}  (\Omega^2-\omega^2) \left(
\Sigma(\Omega^2_{\nu}) - \Sigma(\Omega^2_{\nu}=0) \right) \ . \ee
It is clear now that there is a zero frequency in the RPA spectrum. 

\subsection{Renormalized RPA approximation}

We see from eqs. (\ref{1.25}) and (\ref{1.26}) that if $\kappa_1$ or
$\kappa_2$ are not vanishing, the norm matrix ${\cal N}$ is not diagonal. In
this case, it won't  always be possible 
to write normalization conditions for the
excited states $\vert \nu>$. We therefore impose the supplementary
conditions : $<a_1^{+} a_1^{+}>=<a_1 a_1>=0$ and
$<a_2^{+} a_2^{+}>=<a_2 a_2>=0$, that is $\kappa_1=0$ and $\kappa_2=0$, {\it 
  i.e.}
we neglect pair correlations. 
With these conditions, the matrix ${\cal N}$ stays
 diagonal in the renormalized approximation : 
\be \label{1.60}
{\cal N}=\pmatrix{ {\cal N}_A & 0 \cr  0 & {\cal N}_A \cr } \quad
\hbox{with} \quad {\cal N}_A=\pmatrix{1 & 0 & 0 \cr 0 & 1+\tau_1+\tau_2 & 0
  \cr 0 & 0 & \tau_1 - \tau_2 \cr } \ . \ee
The matrices ${\cal A}$ and ${\cal B}$ 
simplify : 
\be \label{1.61} 
{\cal A}=\pmatrix{\omega & 0 & 0 \cr 0 & (\omega + \Omega)(1 +\tau_1 +\tau_2) 
& 0 \cr 0 & 0 & (\Omega-\omega)(\tau_1-\tau_2) \cr } + {\cal B} \ , \ee
\be \label{1.62}
{\cal B}=\pmatrix{0 & 2h_{12}\Gamma & 2h_{12} \Delta \cr 2 h_{12} \Gamma & 4
  g_{12}\Gamma^2 & 4 g_{12} \Gamma \Delta \cr 2 h_{12} \Delta & 4 g_{12}
  \Gamma \Delta & 4 g_{12} \Delta^2 \cr } \ . \ee
We notice that, contrary to what happen in standard RPA, the operators $a_1
a_2^{+}$ and $a_1^{+}a_2$ don't decouple : the RPA matrix remains 6 by 6.
Its expression is very similar to standard RPA at finite temperature,
$\tau_1$ and $\tau_2$ being the occupation numbers \cite{5}. 

We defined new three-dimensional vectors $\bar {\cal U }_{\nu}$ and $\bar
{\cal V}_{\nu}$ and a new RPA matrix $\bar {\cal R}$ by  : 
\be \label{1.62} 
\pmatrix{ \bar {\cal U }_{\nu} \cr \bar {\cal V}_{\nu} \cr } = {\cal N}^{1/2} 
\pmatrix{  {\cal U }_{\nu} \cr  {\cal V}_{\nu} \cr } \quad \ , \quad 
\bar {\cal R}= {\cal N}^{-1/2} \pmatrix{ {\cal A} & {\cal B} \cr - {\cal B} & 
  -{\cal A} \cr } {\cal N}^{-1/2} \ . \ee
We have : 
\be \label{1.63} 
\bar {\cal R}=\pmatrix{ \bar {\cal A} & \bar {\cal B} \cr - \bar {\cal B} &
  -\bar {\cal A} \cr } \ , \ee
with : $\bar {\cal A}={\cal N}_A^{-1/2} {\cal A} {\cal N}_A^{-1/2}$ and
$\bar {\cal B}={\cal N}_A^{-1/2} {\cal B} {\cal N}_A^{-1/2}$. 
The renormalized RPA equations then write : 
\be \label{1.64} 
\pmatrix{ \bar {\cal A} & \bar {\cal B} \cr -\bar {\cal B} & -\bar {\cal A} 
  \cr } \:   
\pmatrix{ \bar {\cal U}_{\nu} \cr \bar {\cal V}_{\nu} \cr } = \Omega_{\nu}  \: 
\pmatrix{ \bar {\cal U}_{\nu} \cr \bar {\cal V}_{\nu} \cr }  
\ , \ee 
and the RPA frequencies are determined by  : 
\be \label{1.65} 
\det \left[ \left( \bar {\cal A} -\bar {\cal B} \right) \left( \bar {\cal A}
+ \bar {\cal B} \right) - \Omega_{\nu}^2  I \right]=0 \ , \ee
I being the 3 by 3 unity matrix.
The matrices $\bar {\cal A}$ and $ \bar {\cal B}$ are given by : 
\be \label{1.66} 
\bar {\cal A}= \pmatrix{ \omega & 0 & 0 \cr 0 & \omega + \Omega & 0 \cr
0 & 0 & \Omega-\omega \cr } + \bar {\cal B} \ ,  \ee
\be \label{1.67} 
\bar {\cal B}=\pmatrix{0 & 2h_{12} \sqrt{\Gamma} & 2h_{12} \sqrt{ \Delta}
  \cr 2 h_{12} \sqrt{\Gamma} & 4
  g_{12}\Gamma & 4 g_{12} \sqrt{\Gamma \Delta} \cr 2 h_{12} \sqrt{\Delta} &
  4 g_{12}  \sqrt{\Gamma \Delta} & 4 g_{12} \Delta \cr } \ , \ee
where $\Gamma=1 + \tau_1 +\tau_2$ and $\Delta=\tau_1-\tau_2$. 

The norm of the excited states 
\be \label{1.67a}
<\nu \vert \nu>=<0\vert Q_{\nu} Q_{\nu}^{+}\vert 0>=<0 \vert [Q_{\nu},Q_{\nu}^{+}]
\vert 0> \ , \ee
can be chosen to be equal to one. This corresponds to : 
\be \label{1.67b}
\pmatrix{ {\cal U}_{\nu} & {\cal V}_{\nu} \cr } 
\pmatrix{ {\cal N}_{A} & 0 \cr 0 & -{\cal N}_{A}  \cr } 
\pmatrix{ {\cal U}_{\nu} \cr {\cal V}_{\nu} \cr } =1 \ , \ee
or 
\be \label{1.67c}
\pmatrix{ \bar {\cal U}_{\nu} & \bar {\cal V}_{\nu} \cr } 
\pmatrix{ I & 0 \cr 0 &  -I \cr } 
\pmatrix{ \bar {\cal U}_{\nu} \cr \bar {\cal V}_{\nu} \cr } =1 \ , \ee
i.e. : 
\be \label{1.67d}
(\bar {\cal U}_{\nu}^{(1)})^2 + (\bar {\cal U}_{\nu}^{(2)})^2 + 
(\bar {\cal U}_{\nu}^{(3)})^2 - (\bar {\cal V}_{\nu}^{(1)})^2 -
(\bar {\cal V}_{\nu}^{(2)})^2 - (\bar {\cal V}_{\nu}^{(3)})^2 =1 
\ . \ee

By replacing the expressions for $g_{12}$ and $h_{12}$, equation
(\ref{1.65}) can be written in the form : 
\be \ba{ll} \label{1.68} 
(\omega^2-\Omega_{\nu}^2) 
& {\displaystyle \left[ 1 - 4g \left( \frac{\omega
    + \Omega}{\omega \Omega} \frac{\Gamma}{\Omega_{\nu}^2-(\omega+\Omega)^2}
  - \frac{\omega-\Omega}{\omega \Omega}
  \frac{\Delta}{\Omega_{\nu}^2-(\omega-\Omega)^2} \right) \right] } \\ 
& {\displaystyle = -32 g^2 <  X_2>^2 \left[ 
  - \frac{\omega-\Omega}{\omega \Omega}
  \frac{\Delta}{\Omega_{\nu}^2-(\omega-\Omega)^2} + \frac{\omega
    + \Omega}{\omega \Omega} \frac{\Gamma}{\Omega_{\nu}^2-(\omega+\Omega)^2}
  \right] } \ea \ee 
We introduce the quantity : 
\be \label{1.69}
\Sigma_{r} (\Omega_{\nu}^2) = 
\frac{\omega + \Omega}{2\omega \Omega} \frac{1+ \tau_1 + \tau_2}
{\Omega_{\nu}^2-(\omega+\Omega)^2}
  - \frac{\omega-\Omega}{2\omega \Omega}
  \frac{\tau_1-\tau_2}{\Omega_{\nu}^2-(\omega-\Omega)^2} \ , \ee 
which is formally 
similar to the self-energy operator in quantum field theories at
finite temperature. 
Equation (\ref{1.68}) then becomes : 
\be \label{1.70} 
\Omega_{\nu}^2=\omega^2 +64 g^2 < X_2>^2 
\frac{\Sigma_{r}(\Omega_{\nu}^2)}{1  - 8 g \Sigma_{r}(\Omega_{\nu}^2)} \ . \ee
This equation for the RPA frequencies $\Omega_{\nu}$ has the same form as in
standard RPA (eq. (\ref{1.54})) but, in renormalized RPA, $\Sigma_{r}$
contains the densities $\tau_1$ and $\tau_2$ which have to be determined
self-consistently. 

Proceeding in the same way as in standard RPA, let us check the existence of
a zero frequency in renormalized RPA in the case $< X_2> \ne 0$. We
first introduce $I_{\pi}=(2 \tau_1+1) /2 \omega$ and 
$I_{\sigma}=(2 \tau_2+1) /2 \Omega$. The equation for the condensate and the
two generalized gap equations have then the same form as in standard RPA : 
\be \label{1.71}
\mu+ 4g <X_2>^2  + 4 g I_{\pi} + 12 g I_{\sigma}=0 \ee 
\be \label{1.72} 
\omega^2=\mu + 4 g < X_2>^2 + 12 g I_{\pi} + 4 g I_{\sigma} \ee 
\be \label{1.73} 
\Omega^2=\mu + 12 g <X_2>^2 + 4 g I_{\pi} + 12 I_{\sigma} \ee 
and we have again : 
\be \label{1.74} 
\omega^2=8g (I_{\pi}-I_{\sigma}) \ee 
\be \label{1.74a}
\Omega^2=8g < X_2>^2 \ . \ee 
We have : 
\be \label{1.75} 
\Sigma_r(\Omega_{\nu}^2=0) = -\frac{1}{2 \omega \Omega} 
\frac{1+\tau_1 + \tau_2}{(\omega +\Omega)} + 
\frac{1}{2 \omega \Omega} 
\frac{\tau_1 - \tau_2}{(\omega -\Omega)} \ , \ee 
or 
\be 
\Sigma_r(\Omega_{\nu}^2=0)=\frac{I_{\pi}-I_{\sigma}}{\omega^2-\Omega^2} \ .
\ee
Using the two equations (\ref{1.74}) and (\ref{1.74a}), we obtain the same
expression as in standard RPA with $\Sigma$ replaced by $\Sigma_r$ : 
\be \label{1.76}
\Omega_{\nu}^2=\frac{8g}{1
  - 8 g \Sigma_r(\Omega_{\nu}^2)}  (\Omega^2-\omega^2) \left(
\Sigma_r(\Omega^2_{\nu}) - \Sigma_r(\Omega^2_{\nu}=0) \right) \ . \ee

 We have therefore proven the existence of a zero frequency in the RPA
 spectrum in the case where we have a broken symmetry in standard RPA as
 well as in renormalized RPA. This result is very encouraging to apply
 renormalized RPA in the linear sigma model. 

The operator associated to the zero mode is hermitian and can not be
normalized according to (\ref{1.67d}). It is equal to the symmetry generator
$L_3$. The amplitudes $\bar {\cal U}_0$ and $\bar {\cal V}_0$ for the zero
mode are : 
\be 
\bar U_0^{(1)}=-i \sqrt{\frac{\omega}{2}} <X_2> \ee
\be 
\bar U_0^{(2)}=\frac{i}{2} \left(
\sqrt{\frac{\Omega}{\omega}}-\sqrt{\frac{\omega}{\Omega}} \right)
\sqrt{\Gamma} \ee
\be 
\bar U_0^{(3)}=\frac{i}{2} \left(
\sqrt{\frac{\Omega}{\omega}}+\sqrt{\frac{\omega}{\Omega}} \right)
\sqrt{\Delta} \ , \ee
and $\bar {\cal U}_0=\bar {\cal V}_0$. 

In order to obtain a closing of the renormalized RPA eigenvalues problem, one
needs the expressions of the expectation values $\tau_1$ and $\tau_2$ in
terms of the RPA amplitudes. One uses the inversion formula (\ref{1.26a})
and the algebra of the sp(4) group : 
\be \label{1.76a}
a_1^{+}a_1=\frac{1}{2} \left( [a_1a_2,
a_1^{+}a_2^{+}]+[a_1^{+}a_2,a_1a_2^{+}]-1 \right) \ , \ee
\be \label{1.76b}
a_2^{+}a_2=\frac{1}{2} \left( [a_1a_2,
a_1^{+}a_2^{+}]-[a_1^{+}a_2,a_1a_2^{+}]-1 \right) \ . \ee
This provides a set of supplementary equations and allows a closing of the
renormalized RPA eigenvalues problem. 

To be complete, we  finally give the expression of the energy in the
renormalized RPA approximation. From the expression of H with normal ordered
product (\ref{1.12}), we obtain : 
\be \ba{lll} \label{1.77} 
<H>= & {\displaystyle p_{11} (\tau_1 + \kappa_1) + p_{22} (\tau_2 +
\kappa_2) -\omega \kappa_1 -\Omega \kappa_2 } \\
& {\displaystyle + 12 g_{11} (\tau_1 + \kappa_1) (\tau_1 +\kappa_1 +1)
+ 12 g_{22} (\tau_2 + \kappa_2) (\tau_2 +\kappa_2 +1) } \\
& {\displaystyle + 2 g_{12} (2 (\tau_1 + \kappa_1) (\tau_2 + \kappa_2) + \tau_1
  + \kappa_1 + \tau _2 + \kappa_2 ) + E_{HB} } \ea \ . \ee
We have checked that minimization of $<H>$ with respect to $< X_2>,
\omega$ and $\Omega$ gives the equation (\ref{1.39}) for $< X_2>$ and the
two generalized gap equations (\ref{1.40}) and (\ref{1.41}). 

\subsection{Renormalized RPA for the symmetric solution} 

 For the symmetric solution $<X_2>=0$, we have $\omega=\Omega,
 \tau_1=\tau_2$. 
 We still use
 $\kappa_1=\kappa_2=0$. 

The matrix ${\cal N}$ is diagonal with a vanishing matrix ${\cal N}_B$ and : 
\be \label{1.78} 
{\cal N}_A=\pmatrix{ 1 & 0 & 0 \cr 0 & 1 + 2 \tau_1 & 0 \cr 0 & 0 & 0 \cr }
\ee
The renormalised RPA matrix writes : 
\be \label{1.79} 
\bar {\cal A}=\pmatrix{ \omega & 0 & 0 \cr 0 & 2 \omega & 0 \cr 0 & 0 & 0
  \cr} + \bar {\cal B} \quad \hbox{with} \quad  
\bar {\cal B}=\pmatrix{ 0 & 0 & 0 \cr 0 & \frac{2g}{\omega^2}(1+2 \tau_1) &
  0 \cr 0 & 0 & 0 \cr } \ee
For the symmetric solution we therefore obtain a decoupling of the
amplitudes in front of the operators $a_1 a_2^{+}$ and $a_1^{+}a_2$ in
renormalized RPA, contrary to  what happens for the solution with broken
symmetry. 

The RPA frequencies are the solutions of : 
\be \label{1.80} 
\left( \Omega_{\nu}-\omega^2 \right) \left[ \Omega_{\nu}^2 -4
  \omega^2-\frac{8g}{\omega}(1+ 2\tau_1) \right] =0  \ee
and the generalized gap equation writes : 
\be \label{1.81}
\omega^2 = \mu + \frac{8g}{\omega}(2 \tau_1+1) \ee
We have therefore the RPA frequencies : $\Omega_{\nu}^2=\omega^2$ and $
\Omega_{\nu}^2=5 \omega^2-\mu$.

\section{RPA from the time-dependent formalism} 

\setcounter{equation}{0}

In this second part of our paper, we will derive the RPA frequencies from
the linearization of the time-dependent Hartree-Bogoliubov (TDHB) equations.
This approach has been introduced in many-body non-relativistic theories
\cite{11a}. It has also been used in $\lambda \Phi^4$ field theory 
in references
\cite{11}, where small oscillations in the broken phase lead to one and two
meson modes of the theory.  In this formalism, the natural variables are
$<X_i>$ and $<P_i>$ (or $\Phi(\vec x)$ and $\Pi(\vec x)$ in $\lambda \Phi^4$
field theory \cite{11}). 
This formalism is well adapted to dynamical problems. Some
people working on RPA approximations use the formalism with the creation and
annihilation operators, other use the time-dependent variational approach. 
It is interesting to make a close comparison between the two approaches.  

For a two-dimensional system, a Gaussian state at finite temperature can be
described by a vector $\alpha^a$ and a matrix $\Xi^{ab}$, $a=1,2 $: 
\be \label{2.1} 
\alpha^a=\pmatrix{ \bar x^a \cr -i \bar p^a \cr } \quad \ , \quad
\Xi^{ab}=\pmatrix{2 G^{ab} & -i T^{ab} \cr -i T^{ba} & -2 S^{ab} \cr } \ee 
with 
\be \label{2.2}
\bar x^a=<X^a> \quad \ , \quad \bar p^a=<P^a> \ee
\be \label{2.3} 
G^{ab}=<\tilde X^a \tilde X^b> \ee
\be \label{2.4} 
S^{ab}=<\tilde P^a \tilde P^b> \ee 
\be \label{2.5}
T^{ab} = <\tilde X^a \tilde P^b + \tilde P^b \tilde X^a> \ee
with $\tilde X^a=X^a-<X^a>$ and $\tilde P^a=P^a-<P^a>$. The matrices G and S
are symmetric. 

At zero temperature, we have only two independent matrices among G,T and S.
The state can be described by a Gaussian wave function parameterized by 
$\vec{ \bar x}, \vec{ \bar p}, G$ and $\Sigma$ : 
\be \label{2.6} 
\psi(X^1,X^2,t)=\frac{1}{{\cal N}} \exp \left( -<\vec X - \vec{ \bar x} \vert
\frac{1}{4G} + i \Sigma \vert \vec X - \vec{ \bar x} > \right) \exp \left( i <
\vec{ \bar p} \vert \vec X -  \vec{\bar x} > \right) \ , \ee
where the matrix $\Sigma$ is related to the preceeding matrices by : 
\be \label{2.7} 
T^{ab} =2 (G \Sigma + \Sigma G)^{ab} \ee
\be \label{2.8} 
S^{ab}=\frac{1}{4}(G^{-1})^{ab}+4 (\Sigma G \Sigma)^{ab} \ee 

If we work with the operators $a_i,a_i^{+}$, as it is more usual in
many-body problems, one introduces the matrix $\rho$ defined by : 
\be \label{2.9}
(1 + 2 \rho)^{ij}= \pmatrix{ <\tilde a_i \tilde a_j^{+} + \tilde a_j^{+}
  \tilde a_i> & -2 <\tilde a_i \tilde a_j> \cr
  2 <\tilde a_i^{+} \tilde a_j^{+}> & - <\tilde a_i^{+} \tilde a_j +
  \tilde a_j  \tilde a_i^{+}> \cr } \ee
The link with the representation with operators $a_i,a_i^{+}$ and the
representation with $X_i,P_i$ is given by the matrix relation : 
\be \label{2.10}
\Xi'=\frac{1}{2} (1 +\tau)(1 + 2 \rho) (1+\tau) \ee
where 
\be \label{2.11}
\tau=\pmatrix{0 & I \cr -I & 0 \cr} \ , \ee 
I being the 2 by 2 matrix and 
\be \label{2.12}
\Xi'=\pmatrix{ 2 \sqrt{\omega_i} <\tilde X_i \tilde X_j> \sqrt{\omega_j} & 
-i \sqrt{\omega_i} < \tilde X_i \tilde P_j+ \tilde P_j \tilde X_i>
\frac{1}{\sqrt{\omega_j}} \cr 
-i \sqrt{\omega_j} < \tilde X_j \tilde P_i+ \tilde P_i \tilde X_j>
\frac{1}{\sqrt{\omega_i}} & -2 \frac{1}{\sqrt{\omega_i}} <\tilde P_i \tilde
P_j> \frac{1}{\sqrt{\omega_j}} \cr } \ee
with $\omega_1=\omega$ and $\omega_2=\Omega$ (see eqs.
(\ref{1.2}),(\ref{1.3})).  

\subsection{The time-dependent Hartree-Bogoliubov equations} 

In this part of the paper, we will use the following notation for the
Hamiltonian of 
two-dimensional quantum anharmonic oscillator : 
\be \label{2.14} 
H=\frac{P_1^2}{2m} +\frac{P_2^2}{2m} + g (X_1^2 +X_2^2-a^2)^2  \ .
\ee

The time-dependent variational Hartree-Bogoliubov equations are obtained
at zero temperature from the minimization of : 
\be \label{2.14a}
S=\int dt \: < \psi \vert i \partial_t-H \vert \psi > \ , \ee
where $|\psi>$ is the state corresponding to the variational wave function
(\ref{2.6}). At finite temperature, the TDHB equations are obtained by
minimizing \cite{11b} 
\be \label{2.14b} 
{\cal Z}({\cal D}(t))=tr({\cal D}(t_1))-\int_{t_0}^{t_1} dt \: tr \left(
\frac{d {\cal D}(t)}{dt} + i [H, {\cal D}(t)] \right) \ , \ee
where ${\cal D}(t)$ is a variational density matrix, chosen to be a Gaussian
and therefore characterized by $\alpha$ and $\Xi$. 

The TDHB equations can be written
in the following compact form \cite{12} :
\be \label{2.15} 
i \dot \alpha=\tau w \ , \ee
\be \label{2.16} 
i \dot \Xi = - \left[ \left( \Xi+\tau \right) {\cal H} \left( \Xi - \tau
\right) - \left( \Xi-\tau \right) {\cal H} \left( \Xi + \tau
\right)  \right] \ , \ee 
or 
\be \label{2.17}
i \dot \Xi=2 \left[ \Xi \: {\cal H} \: \tau - \tau \: {\cal H} \: 
\Xi \right] \ . \ee
The vector $w$ and the matrix ${\cal H}$ are defined by : 
\be \label{2.18} 
\delta <H> =\tilde w_i^a \delta \alpha_i^a -\frac{1}{2} tr \left( {\cal
  H}_{ij}^{ab} \delta \Xi_{ji}^{ba} \right) \ . \ee 
For the Hamiltonian (\ref{2.14}), we have : 
\be  \ba{llll} \label{2.19}
<H>= & {\displaystyle \frac{1}{2m} (p_1^2+p_2^2+tr S) } \\
& {\displaystyle + g G_{11} (6 \bar x_1^2 +2 \bar x_2^2 -2 a^2+3 G_{11} +
  G_{22} ) } \\
& {\displaystyle + g G_{22} (2 \bar x_1^2 +6 \bar x_2^2 -2 a^2+ G_{11} +
  3 G_{22} ) } \\
& {\displaystyle + g G_{12} (8 \bar x_1 \bar x_2 + 4 G_{12}) + g (\bar x_1^2
  + \bar x_2^2 -a^2)^2 } \ea \ . \ee
The vector $w$ and the matrix ${\cal H}$ are given by : 
\be \label{2.20} 
w_2^1=\frac{i}{m}\bar p_1 \quad \ , \quad  w_2^2=\frac{i}{m}\bar p_2 \ee 
\be \label{2.21}
w_1^1=4 g \bar x_1 \left( \bar x_1^2 + \bar x_2^2 -a^2 \right) +
12 g \bar x_1
G^{11} +4 g \bar x_1 G^{22} +8 g \bar x_2 G^{12} \ee 
\be \label{2.22}
w_1^2=4 g \bar x_2 \left( \bar x_1^2 + \bar x_2^2 -a^2 \right) +
12 g \bar x_2
G^{22} +4 g \bar x_2 G^{22} +8 g \bar x_1 G^{12} \ee 

\be \label{2.23} 
{\cal H}_{11}^{11}=-6g \bar x_1^2 -2g \bar x_2^2 -6g G^{11} -2g G^{22}
+2g a^2 \ee
\be \label{2.24} 
{\cal H}_{11}^{12}=-4g \bar x_1 \bar x_2 -4g G^{12} = {\cal H}_{11}^{21} \ee
\be \label{2.25} 
{\cal H}_{11}^{22}=-6g \bar x_2^2 -2g \bar x_1^2 -6g G^{22} -2g G^{11}
+2g a^2 \ee
\be \label{2.26} 
{\cal H}_{22}^{ab}\equiv \frac{\delta <H>}{\delta S^{ba}} = \frac{1}{2m}
\delta^{ab} \ee
\be \label{2.27} 
{\cal H}_{12}^{ab}\equiv 2 i \frac{\delta <H>}{\delta T^{ab}}=0 \quad \ ,
\quad {\cal H}_{21}^{ab}\equiv 2 i \frac{\delta <H>}{\delta T^{ba}}=0 \ee

\subsection{Static solution of the TDHB equations} 

The static solution of the TDHB equations (\ref{2.15}) and (\ref{2.17}) is
given by : 
\be \label{2.28} 
\bar w=0 \ee
\be \label{2.29} 
\bar \Xi \: {\cal H}(\bar \alpha, \bar \Xi) \: \tau - \tau \: 
{\cal H}(\bar \alpha, \bar
\Xi) \: \bar \Xi =0 \ee
Using the rotational invariance we can choose $\bar x^1=0$ and from $\bar
w=0$ we deduce also $G^{12}=0$ (to simplify the notations we don't use the
bar on the matrix elements of $\Xi$ for the HFB ground state). From $\bar
w=0$, we deduce the existence of two solutions. One solution is symmetric
with $\bar x_2=0$. The other solution shows a broken symmetry with $\bar x_2
\ne 0$ and given by : 
\be \label{2.30} 
\bar x_2^2=a^2-G^{11}-3 G^{22} \ee 
>From eq. (\ref{2.29}), we deduce : 
\be \label{2.31} 
T^{ab}=0
\ee
\be \label{2.32}
S^{12}=0 \ee 
\be \label{2.33} 
S^{11} =4m g G^{11}(\bar x_2^2-a^2 +3 G^{11} + G^{22}) \ee 
\be \label{2.34} 
S^{22} =4m g G^{22}(3\bar x_2^2-a^2 + G^{11} + G^{22}) \ee 
For the symmetric solution, we have : 
\be \label{2.35} 
S^{11}=S^{22}=4mg (4 G^{11}-a^2) \ . \ee
For the solution with broken symmetry, 
\be \label{2.36} 
S^{11} =8gm(G^{11}-G^{22})G^{11} \ee 
\be \label{2.37} 
S^{22}=8gm G^{22} \bar x_2^2 \ee 

At zero temperature, we have for the static solution :
$T^{ab}=0$ and $S^{11}=1/4 G^{11}, S^{22}=1/4 G^{22}$. Equations
(\ref{2.33}) and (\ref{2.34}) give the two gap equations : 
\be \label {2.38} 
\frac{1}{8m}G_{11}^{-2}=2 g (\bar x_2^2-a^2 + 3 G_{11} + G_{22}) \ee
\be \label {2.39} 
\frac{1}{8m}G_{22}^{-2}=2 g (3\bar x_2^2-a^2 +  G_{11} + G_{22}) \ee
By using the identification $G^{11}=1/2\omega$ and $G^{22}=1/2\Omega$, 
the previous equations are identical to the mean-field gap
equations obtained in the first part of the paper (\ref{1.16}) and 
(\ref{1.17}). 

At zero temperature, the symmetric phase is characterized by : 
\be 
\bar x_2=0 \ee 
\be \label{2.40} 
G_{11}=G_{22} \quad \ , \quad \frac{1}{8m} G_{11}^{-2}=2 g (4G_{11} -a^2) \
. \ee
The solution with broken symmetry at zero temperature is characterized by : 
\be \label{2.41} 
\bar x_2^2 =a^2 -G_{11} -3 G_{22} \ee
\be \label{2.42} 
\frac{1}{32m} G_{11}^{-2}=g (G_{11}-G_{22}) \ee
\be \label{2.43}
\frac{1}{32m} G_{22}^{-2}=g \bar x_2^2 \ee 

\subsection{Small oscillations around the static solution} 

 The linearization of the TDHB equations (\ref{2.15}) and (\ref{2.17}) writes 
\be \label{2.44}
i \delta \dot \alpha =\delta w \ee 
\be \label{2.45} 
i \delta \dot \Xi=2 \left[ 
\delta \Xi \: \bar {\cal H} \: \tau -\tau \: \bar {\cal H} \: \delta \Xi + 
\bar \Xi
\: \delta {\cal H} \: \tau -\tau \: \delta {\cal H} \: \bar \Xi \right] \ee
where $\bar {\cal H}$ is the matrix ${\cal H}$ evaluated for the HFB static
solution $\bar \alpha, \bar \Xi$. 

Let us write more explicitly the linearization of the TDHB equations around
the static HFB solution with {\it broken symmetry} characterized by
$\bar x_1=0,\bar x_2 \ne 0, G^{11},G^{22},S^{11},S^{22},$$ \bar p_1=\bar p_2=0,
G^{12}=S^{12}=T^{ab}=0$. We obtain a differential system with 14 variables :
$\delta x_1, \delta p_1, \delta x_2, \delta p_2, \delta G^{11}, \delta
G^{12}, \delta G^{22}, \delta T^{11}, \delta T^{12}, \delta T^{21}, \delta
T^{22}, \delta S^{11}, \delta S^{12}, \delta S^{22}$ : 
\be 
\delta \dot x_1=\frac{1}{m} \delta p_1 \ee
\be 
\delta \dot p_1 =-8 g (G^{11}-G^{22}) \delta x_1 -8 g \bar x_2 \delta G^{12}
\ee
\be 
\delta \dot x_2=\frac{1}{m} \delta p_2 \ee
\be 
\delta \dot p_2 =-8 g \bar x_2^2-12 g \bar x_2 \delta G^{22} \ee 

\be 
\delta \dot G^{11}=\frac{1}{m} \delta T^{11} \ee 
\be 
\delta \dot G^{12}=\frac{1}{2m} \delta T^{12} + \frac{1}{2m} \delta T^{21} 
\ee 
\be 
\delta \dot G^{22}=\frac{1}{m} \delta T^{22} \ee 

\be 
\delta \dot T^{11}=-16g(G^{11}-G^{22}) \delta G^{11} +\frac{2}{m} \delta
S^{11} \ee 
\be 
\delta \dot T^{12}=-16 g G^{11} \bar x_2 \delta x_1 -16 g (\bar x_2^2 +
G^{11}) \delta G^{12} + \frac{2}{m} \delta S^{12} \ee
\be 
\delta \dot T^{21}=-16 g G^{22} \bar x_2 \delta x_1 -16 g G^{11} \delta
G^{12} + \frac{2}{m} \delta S^{12} \ee 
\be 
\delta \dot T^{22} =-16 g \bar x_2^2 \delta G^{22} + \frac{2}{m} \delta
S^{22} \ee

\be
\delta \dot S^{11}=-8g (G^{11}-G^{22}) \delta T^{11} \ee
\be 
\delta \dot S^{12}=-4g (G^{11}-G^{22}) \delta T^{12}-4 g \bar x_2^2 \delta
T^{21} \ee
\be 
\delta \dot S^{22}=-8 g \bar x_2^2 \delta T^{22} \ee 
To write these equations, we have used the mean-field equations for the
solution with broken symmetry. 

>From this differential system of first order, we obtain a 14 by 14 RPA
matrix. Coming back to the operator representation, with the help of
relation (\ref{2.10}) and its inverse, this corresponds to the 14 operators 
we have to include in the most general form of 
the excitation operator $Q_{\nu}^{+}$ of the first part of this paper if we
keep only bilinear operators :
$a_1, a_1^{+},a_2,a_2^{+}, a_1a_1,a_1^{+}a_1^{+}
,a_2a_2, a_2^{+}a_2^{+}, a_1a_2,
a_1^{+}a_2^{+},a^{+}_1a_2,a_1a^{+}_2$ and $a^{+}_1a_1,a^{+}_2a_2$. 
The last two operators appear only at finite temperature (For the derivation
of RPA equations at finite temperature, see reference \cite{13}). 

The operator $L_3$ allows to separate these operators in two sectors :
the six operators $a_1, a_1^{+},  a_1a_2,
a_1^{+}a_2^{+},a^{+}_1a_2,a_1a^{+}_2$ corresponding to the ``pion'' sector
and the 8 operators
$a_2,a_2^{+}, a_1a_1,a_1^{+}a_1^{+},a_2a_2, a_2^{+}a_2^{+}, a_1a_2,
a_1^{+}a_2^{+},a^{+}_1a_2,a_1a^{+}_2,a^{+}_1a_1,a^{+}_2a_2$ corresponding to
the ``sigma''sector. By using again the correspondence between the $X_i,P_i$
representation and the $a^{+}_i,a_i$ representation, we deduce that the
variables in our first-order differential system which correspond to the
``pion'' are : $\delta \bar x_1, \delta \bar p_1, \delta G^{12}, 
\delta T^{12}, \delta T^{21}, \delta S^{12}$, the
other corresponding to the ``sigma'' sector. 
We check indeed that the RPA matrix, 
which we call ${\cal R}$, can be written in two blocks, 
one 6 by 6 corresponding to the
``pion'' and one 8 by 8 corresponding to the ``sigma'', the two sectors
being disconnected. 
We can write the first-order differential system in the form : 
\be \label{2.46}  
\delta \dot X = {\cal R} \delta X \ , \ee
where $\delta \tilde X= (\delta \bar x_1, \delta \bar p_1, \delta G^{12},
\delta T^{12}, \delta T^{21}, \delta S^{12}, \delta \bar x_2, \delta \bar p_2, 
\delta
G^{11}, \delta G^{22}, \delta T^{11}, \delta T^{22}, \delta S^{11}, \delta
S^{22})$ and 
\be \label{2.47}
{\cal R}=\pmatrix{ {\cal M}_{\pi} & 0 \cr 0 & {\cal M}_{\sigma} \cr } \ee
The 6 by 6 RPA matrix ${\cal M}_{\pi}$ for the ``pion'' sector is equal to : 
\be \label{2.47} 
{\cal M}_{\pi}=\pmatrix{0 & \frac{1}{m} & 0 & 0 & 0 & 0 \cr 
-8 g (G^{11}-G^{22}) & 0 & -8g \bar x_2 & 0 & 0 & 0 \cr 
0 & 0 & 0 & \frac{1}{2m} & \frac{1}{2m} & 0 \cr 
-16 g G^{11}\bar x_2 & 0 & -16 g (\bar x_2 ^2 + G^{11}) & 0 & 0 &
\frac{2}{m} \cr 
-16 g G^{22}\bar x_2 & 0 & -16 g  G^{11} & 0 & 0 &
\frac{2}{m} \cr 
0 & 0 & 0 & -4 g (G^{11}-G^{22}) & -4g \bar x_2^2 & 0 \cr }
 \ee
and the 8 by 8 RPA matrix ${\cal M}_{\sigma}$ is equal to : 
\be \label{2.48} 
{\cal M}_{\sigma}= \pmatrix{ 0 & \frac{1}{m} & 0 & 0 & 0 & 0 & 0 & 0 \cr 
-8g \bar x_2^2 & 0 & 0 & -12g \bar x_2 & 0 & 0 & 0 & 0 \cr 
0 & 0 & 0 & 0 & \frac{1}{m} & 0 & 0 & 0 \cr 
0 & 0 & 0 & 0 & 0 & \frac{1}{m} & 0 & 0 \cr 
0 & 0 & -16 g (G^{11}-G^{22}) & 0 & 0 & 0 & \frac{2}{m} & 0 \cr  
0 & 0 & 0 & -16 g \bar x_2^2 & 0 & 0 & 0 & \frac{2}{m} \cr 
0 & 0 & 0 & 0 & -8 g (G^{11}-G^{22}) & 0 & 0 & 0 \cr 
0 & 0 & 0 & 0 & 0 & -8 g \bar x_2^2 & 0 & 0 \cr }
\ee
We have $\det {\cal M}_{\pi}=0$. However, before to conclude about the
existence of a zero mode associated to the spontaneously breakdown of the
rotational symmetry,  we
have to eliminate the spurious modes corresponding to invariants of the TDHB
evolution.  At zero temperature, 
this will allow to
reduce the ${\cal M}_{\pi}$ matrix to a 4 by 4 matrix. This is   also the
dimension we have found 
in the first section in standard RPA. 

At zero temperature, we have the following condition to be satisfied to have
a pure state : 
\be \label{2.49}
\rho (\rho + I)=0 \ee 
or 
\be \label{2.50}
-\Xi' \tau \Xi'=\tau \ , \ee
where the matrices $\rho$ and $\Xi'$ are given by eqs. (\ref{2.9}) and
(\ref{2.12}) and I is the 2 by 2 unity matrix. We linearize eq. (\ref{2.50})
around the HFB static solution. This gives the following conditions : 
\be \label{2.51}
\sum_k  G^{ik} \delta T^{jk} -\delta T^{ik}  G^{kj}=0 \ee
\be \label{2.52} 
\sum_k \delta T^{ki}  S^{kj} -  S^{ik} \delta T^{kj}=0 \ee
\be \label{2.53}
\sum_k G^{ik} \delta S^{kj} + \delta G^{ik} S^{kj} =0 \ee
Equations (\ref{2.51}) and (\ref{2.52}) give respectively : 
\be \label{2.54}
G^{11} \delta T^{21}-\delta T^{12} G^{22} =0 \ee
\be \label{2.55}
S^{22} \delta T^{21}-\delta T^{12} S^{11} =0 \ee
At the minimum at zero temperature, we have : $S^{11}=1/4 G^{11}$ and
$S^{22}=1/4 G^{22}$. The two previous conditions are therefore equivalent. 
>From eq. (\ref{2.53}), we obtain : 
\be \label{2.56} 
\delta G^{12} +4 G^{11} G^{22} \delta S^{12}=0 \ee
\be \label{2.57} 
\delta G^{11} + 4 (G^{11})^2 \delta S^{11}=0 \ee
\be \label{2.58} 
\delta G^{22} + 4 (G^{22})^2 \delta S^{22} =0 \ee  
The first condition is for the ``pion'' sector and the last two conditions
are for the ``sigma''sector. 

Coming back to the operator representation, we have : 
\be \label{2.58a}
2 \sqrt{\omega \Omega} <\tilde X_1 \tilde X_2> + \frac{2}{\sqrt{\omega
    \Omega}} < \tilde P_1 \tilde P_2 > =2 ( <a_1 a_2^{+}> + <a_1^{+} a_2>) \ee 
\be \label{2.58b}
2 \sqrt{\frac{\omega}{\Omega}} <\tilde X_1 \tilde P_2 + \tilde P_2 \tilde
X_1> -2 \sqrt{\frac{\Omega}{\omega}}<\tilde X_2 \tilde P_1 + \tilde P_1
\tilde X_2> =4 ( <a_1 a_2^{+}> - <a_1^{+}a_2>) \ee
By using $G^{12}=<\tilde X_1 \tilde X_2>, S^{12}=< \tilde P_1 \tilde P_2>,
T^{12}=2<\tilde X_1 \tilde P_2>,T^{21}=2 <\tilde X_2 \tilde P_1>$, and
$G^{11}=1/\omega, G^{22}=1/\Omega$ for the static HFB solution, we see that
conditions (\ref{2.54}) and (\ref{2.56}) correspond to the decoupling of the
operators $a_1 a_2^{+}, a_1^{+}a_2$ we have found in the first section in
standard RPA. 

For the pion sector, we will therefore consider the  following new variables : 
\be \ba{lll} \label{2.59}
\delta \tilde Y= ( 
& {\displaystyle \delta x_1, \delta p_1, \delta C=-4 G^{11}G^{22} \delta
G^{12}+ \delta S^{12}, \delta D = G^{11} \delta T^{12} + G^{22} \delta
T^{21}, } \\ 
& {\displaystyle \delta E=\delta G^{12}+G^{22}\delta T^{12}-G^{11} \delta T^{21} + 4
G^{11} G^{22} \delta S^{12} ,} \\
& {\displaystyle \delta F=\delta G^{12}-G^{22}\delta T^{12}+
G^{11} \delta T^{21} + 4 G^{11} G^{22} \delta S^{12} ) } \ea  \ee
We have : 
\be \label{2.60} 
\delta \dot Y = {\cal M}'_{\pi} \: \delta Y \ee
>From the condition to remain in the variational space of pure Gaussian states
we have : $\delta E=0, \delta F=0$.
We therefore consider in ${\cal M}_{\pi}'$ the 4 by 4 matrix corresponding
to the four coordinates $\delta x_1,\delta p_1, \delta C, \delta D$ (we
notice that there are non-vanishing matrix elements in the fifth and sixth
columns of ${\cal M}_{\pi}'$) : 
\be \label{2.61} 
{\cal M}''_{\pi}=\pmatrix{0 & \frac{1}{m} & 0 & 0 \cr 
-8 g (G^{11}-G^{22}) & 0 & \frac{32 G^{11} G^{22} g \bar x_2}{1 + 16 (G^{11}
  G^{22})^2} & 0 \cr 0 & 0 & 0 & N_1 \cr 
-16 ((G^{11})^2 +(G^{22})^2) g \bar x_2 & 0 & N_2 & 0 \cr } \ee 
where 
\be \label{2.62} 
N_1=-\frac{2}{m((G^{11})^2+(G^{22})^2)} \left[ G^{11} G^{22} (G^{22}-2
gm)+(G^{11})^2(G^{22}+2gm)+2G^{22}g \bar x_2^2m \right] \ee
\be \label{2.63}
N_2= -\frac{2}{m(1+(G^{11}G^{22})^2)} \left[ G^{11} +G^{22} +32
(G^{11})^3G^{22}gm +32 G^{{11}^2}G^{22} gm (G^{22}+\bar x_2^2) \right] \ee
We then calculate $\det {\cal M}_{\pi}''$ and, by using the two gap equations 
(\ref{2.42}) and (\ref{2.43}) for the solution with broken symmetry, we
obtain : 
\be \label{2.64} 
\det {\cal M}_{\pi}''=0 \ee
At zero temperature, we have therefore checked the existence of a zero
frequency for the solution with broken symmetry in the standard RPA. 

At finite temperature, the condition (\ref{2.49}) becomes : 
\be \label{2.65} 
\rho (\rho+ I) =\frac{1}{4} (C-1) I \ee
where C is called the Heisenberg invariant and is a quantity conserved by
the TDHB evolution : $\dot C=0$. In our O(2) model, C has two indices :
$C^{ab}, a,b=1,2$. The Heisenberg invariant is equal to : 
\be \label{2.66}
C^{ab}=\sum_c 4 <\tilde X^a \tilde X^c> <\tilde P^c \tilde P^b>-
<\tilde X^a \tilde P^c + \tilde P^c \tilde X^a> <\tilde X^c \tilde P^b +
\tilde P^b \tilde X^c> \ee
For the static solution at finite temperature, we have $T^{ab}=0$ and : 
\be
4 G^{11} S^{11}=C^{11} \: \: \ , \: \: 4 G^{22} S^{22}=C^{22} \ . \ee
$C^{11}$ and $C^{22}$ are related to the occupation numbers for the bosons 1
and 2 according to : $2n_a+1=\sqrt{C^{aa}}$.
Similarly to the zero temperature case, we linearize the condition
(\ref{2.65}) around the static solution characterized by $G^{11}, G^{22},
S^{11}$ and $S^{22}$. It is then convenient to introduce new variables in
the pion sector and we check the existence of the zero mode. 

\section{Conclusion}

 In this paper, we have applied 
 the random phase approximation and its extension
 called renormalized RPA to the quantum anharmonic oscillator with an O(2)
 symmetry. The expression for the RPA matrix in renormalized RPA is formally 
 very
 similar to the RPA matrix appearing in standard RPA at finite temperature. 
We focused on the existence of a zero mode among the RPA frequencies in the
case where the ground state has a broken symmetry. This result is
encouraging to apply renormalized RPA in the linear sigma model. 
We have compared also the approach with the creation and annihilation operators
with the time-dependent approach and we identify the variables corresponding
to the ``pion'' sector and those corresponding to the ``sigma'' sector. 
The numerical resolution of the self-consistent renormalized RPA equations
will be the subject of a next paper, where Hartree-Bogoliubov mean-field
results, standard RPA results and renormalized RPA results will be compared
to exact numerical results for the vacuum energy and the energy of the first
excited states in the cases of the symmetric solution and the solution with
broken symmetry. 

\vspace*{1cm} 

{\bf Aknowledgements}

I am grateful to  Peter Schuck for very helpful discussions.

\vspace*{1cm}







\end{document}